\documentclass[12pt]{article}
\usepackage{graphicx}
\usepackage{amssymb}
\usepackage{bm}
\usepackage{setspace}

\begin{document}

\title{Circular and helical equilibrium solutions of inhomogeneous rods}

\author{Alexandre F. da Fonseca, C. P. Malta}

\date{}

\maketitle

\begin{center}
{\small{Instituto de F\'{\i}sica, Universidade de S\~ao Paulo,
USP\\ Rua do Mat\~ao, Travessa R 187, Cidade Universit\'aria,
05508-900, S\~ao Paulo, Brazil}}
\end{center}

\begin{abstract}

Real filaments are not perfectly homogeneous. Most of them have
various materials composition and shapes making their stiffnesses not
constant along the arclength. We investigate the existence of circular
and helical equilibrium solutions of an intrinsically straight rod
with varying bending and twisting stiffnesses, within the framework of
the Kirchhoff model. The planar ring equilibrium solution only exists
for a rod with a given form of variation of the bending stiffness. We
show that the well known {\it circular helix} is not an equilibrium
solution of the static Kirchhoff equations for a rod with non constant
bending stiffness. Our results may provide an explanation for the
variation of the curvature seen in small closed DNAs immersed in a
solution containing Zn$^{2+}$, and in the DNA wrapped around a
nucleosome.

\end{abstract}

\noindent {\small{\textbf{Keywords:} Kirchhoff rod model, inhomogeneous rod, 
planar ring, generalized helices, planar elasticae, closed DNA}}




\thispagestyle{empty}

\begin{doublespace}

\section{Introduction}  

Static and dynamics of one-dimensional structures have been the
subject of intense research in Engineering, Physics and Biology. Such
filamentary objects are present in both nature and man-made devices.
The stability of sub-oceanic cables~\cite{zajac,coyne}, the
tridimensional structure of biomolecules~\cite{tamar} and bacterial
fibers~\cite{wolge}, and the mechanical properties of
nanosprings~\cite{mc,fonseca3,huang,fonseca4} are few of many
interesting problems investigated within elastic rod models.

In most cases, the rod is considered as having constant or uniform
stiffness, but non-uniform rods have been considered in the
literature. It has been shown that nonhomogeneous Kirchhoff rods may
present spatial chaos~\cite{holmes,davies}. Domokos and Holmes studied
buckled states of non-uniform elasticae~\cite{domokos1}. Deviations
from the helical structure of rods due to periodic variation of the
Young's modulus were verified numerically by da Fonseca, Malta and de
Aguiar~\cite{fonseca0}. Homogeneous and nonhomogeneous rods subject to
given boundary conditions were studied by da Fonseca and de Aguiar
in~\cite{fonseca1}. The effects of a nonhomogeneous mass distribution
in the dynamics of unstable closed rods have been analyzed by Fonseca
and de Aguiar~\cite{fonseca2}. Goriely and McMillen~\cite{alain2}
studied the dynamics of cracking whips~\cite{whip}, Kashimoto and
Shiraishi~\cite{moto} studied twisting waves in inhomogeneous rods,
and Neuringer and Elishakoff~\cite{neuri} studied natural frequencies
of inhomogeneous rods.

It is well known that straight, circular and helical rods are
equilibrium solutions of the Kirchhoff model~\cite{kirch,dill,nize}.
These solutions have been used to study the stability and buckling
phenomena of loop elasticae~\cite{zajac,gori1}, the {\it twist to
writhe} conversion phenomenon in twisted straight rods~\cite{goriely},
and the stability of DNA configurations with or without
self-contact~\cite{col1}, to name some examples.

Our aim here is to investigate the existence of circular and helical
equilibrium solutions for an intrinsically straight, inextensible,
unshearable, and inhomogeneous rod with circular cross-section, within
the Kirchhoff rod model. 

Establishing the stability (even in the case of simple equilibrium
configurations) of a rod is a non-trivial task (see, for example,
Refs. \cite{gori1,goriely,col1,tabor2,patrick,nize1,nadia}). Since the
study of the stability analysis departs from the knowledge of
equilibrium solutions, our work is the starting point for the study of
the stability of equilibrium solutions of inhomogeneous rods.

The inhomogeneity of the rod is considered through the bending and
twisting coefficients varying along its arclength $s$, $B(s)$ and
$C(s)$, respectively. We shall derive a set of non-linear differential
equations for the curvature, $k_F(s)$, and the torsion, $\tau_F(s)$,
of the centerline of an inhomogeneous rod, and then impose the
necessary conditions to finding circular/helical solutions. Circular
(ring) solutions have $k_F=$ Constant and $\tau_F=0$, while helical
solutions have the curvature and torsion of the rod centerline
satisfying the {\it Lancret's theorem}: $k_F(s)/\tau_F(s)$ is
constant~\cite{manfredo,dirk}. The well known {\it circular helix} is
the particular case where both $k_F$ and $\tau_F$ are constant.

The {\it fundamental theorem} for space curves~\cite{dirk} states that
the curvature and the torsion completely determine a space curve, but
for its position in space. This is the reason for our choice of
working with the curvature, $k_F(s)$, and the torsion, $\tau_F(s)$, of
the rod centerline. We shall show that the solutions for $k_F(s)$ and
$\tau_F(s)$ depend only on the bending coefficient, $B(s)$, an
expected result since the centerline of the rod does not depend on the
twisting coefficient (see for instance, Neukirch and
Henderson~\cite{neuk}).

In Sec. II we present the static Kirchhoff equations for an
intrinsically straight rod with circular cross section and varying
stiffness, and derive the non-linear differential equations for the
curvature and torsion of the rod. In Sec. III we analyse the cases of
null torsion (straight and planar rods). We show that the ring
solution of the static Kirchhoff equations only exists for a
particular form of variation of the bending stiffness. In the case of
twisted rods with varying bending stiffness, this ring solution is the
only possible planar solution of the Kirchhoff equations. In the case
of non-twisted rods, there exist other types of equilibrium planar
solutions that depend on the bending stiffness. In Sec. IV we use the
{\it Lancret's theorem} for obtaining generalized helical solutions of
the static Kirchhoff equations. We show that the {\it circular helix}
is not an equilibrium solution of a rod that has non constant bending
stiffness. We obtain generalized helical solutions that depend on the
form of variation of the bending stiffness. As illustration, we
compare the helical solution of a homogeneous rod with two types of
helical solutions related to simple cases of inhomogeneous rods: (i)
bending coefficient varying linearly, and (ii) bending varying
periodically along the rod. In Sec. V we discuss the main results and
possible applications to DNA.

\section{The static Kirchhoff equations} 

The statics and dynamics of long, thin, inextensible and unshearable
elastic rods are described by the Kirchhoff rod model. In this model,
the rod is divided in segments of infinitesimal thickness to which the
Newton's second law for the linear and angular momenta are applied. We
derive a set of partial differential equations (PDE) for the averaged
forces and torques on each cross section and for a triad of vectors
describing the shape of the rod. The set of PDE are completed with a
linear constitutive relation between torque and twist.

The central axis of the rod, hereafter called centerline, is
represented by a space curve ${\mathbf{x}}$ parametrized by the
arclength $s$. We commonly describe a physical filament using a local
basis, $\{{\mathbf{d}}_1,{\mathbf{d}}_2,{\mathbf{d}}_3\}$, which
permits taking into account the twist deformation of the
filament. This local basis is defined such that ${\mathbf{d}}_3$ is
the vector tangent to the centerline of the rod
(${\mathbf{d}}_3={\mathbf{t}}\equiv d{\mathbf{x}}/ds$), and
${\mathbf{d}}_1$ and ${\mathbf{d}}_2$ lie on the cross section
plane. The local basis is related to the {\it Frenet} frame
$\{{\mathbf{n}},{\mathbf{b}},{\mathbf{t}}\}$ of the centerline through
\begin{equation}
\label{def_xi}
({\mathbf{d}}_1 \; {\mathbf{d}}_2 \; {\mathbf{d}}_3)=
({\mathbf{n}} \; {\mathbf{b}} \; {\mathbf{t}})
\left(
\begin{array}{ccc}
\cos\xi & -\sin\xi & 0  \\
\sin\xi & \cos\xi & 0  \\
0 & 0 & 1
\end{array}
\right)  \; ,
\end{equation}
where the angle $\xi$ is the amount of twisting of the local basis
with respect to ${\mathbf{t}}$.

We are interested in the equilibrium solutions of the Kirchhoff model,
so our study departs from the static Kirchhoff equations~\cite{foot2}.
For intrinsically straight isotropic rods, these equations are:
\begin{eqnarray}
&&\mathbf{F}'=0 \; , \label{kir1_a} \\
&&\mathbf{M}'=\mathbf{F}\times\mathbf{d}_{3} \; , \label{kir1_b} \\
&&\mathbf{M}=B(s)\,k_1\,\mathbf{d}_1+B(s)\,k_2\,\mathbf{d}_2+
C(s)\,k_3\,\mathbf{d}_3 \; , \label{kir1_c}
\end{eqnarray}
where the following scaled variables were introduced:
\begin{equation}
\label{convert}
s\rightarrow sL \; , \; \; \; \mathbf{F}\rightarrow 
\mathbf{F}\frac{EI}{L^2} \; , \; \; \; 
\mathbf{M}\rightarrow \mathbf{M}\frac{EI}{L} \; \; \; \mbox{and} \; \; \;
k_i\rightarrow k_i\frac{1}{L} \;  , i=1,2,3 \; .
\end{equation}
the vectors $\mathbf{F}$ and $\mathbf{M}$ are the resultant force and
the corresponding moment with respect to the centerline of the rod,
respectively, at a given cross section. The prime $'$ denotes
differentiation with respect to $s$. $k_i$ are the components of the
twist vector, $\mathbf{k}$, that controls the variations of the
director basis along the rod through the relation
\begin{equation}
\label{dis}
{\mathbf{d}}'_i={\mathbf{k}}\times{\mathbf{d}}_i \; , \; \; i=1,2,3 \; . 
\end{equation}
$k_1$ and $k_2$ are related to the curvature of the centerline of the
rod $(k_F=\sqrt{k^{2}_{1}+k^{2}_{2}})$ and $k_3$ is the twist
density. $B(s)$ and $C(s)$ are the bending and twisting coefficients
of the rod, respectively. Writing the force $\mathbf{F}$ in the
director basis,
\begin{equation}
\label{F}
{\bf{F}}=f_1{\bf{d}}_1+f_2{\bf{d}}_2+f_3{\bf{d}}_3 \; , 
\end{equation}
the equations (\ref{kir1_a}--\ref{kir1_c}) give the following
differential equations for the components of the force and twist
vector:
\begin{eqnarray}
f'_1-f_2\, k_3+f_3\, k_2=0 \; , \label{a}  \\
f'_2+f_1\, k_3-f_3\, k_1=0 \; , \label{b}  \\
f'_3-f_1\, k_2+f_2\, k_1=0 \; , \label{c}  \\
(B(s)\, k_1)'+(C(s)-B(s))\, k_2\, k_3-f_2=0 \; , \label{d} \\
(B(s)\, k_2)'-(C(s)-B(s))\, k_1\, k_3+f_1=0 \; , \label{e} \\
(C(s)\ k_3)'=0 \; . \label{f} 
\end{eqnarray}
The equation (\ref{f}) shows that the component $M_3=C(s)\,k_3$ of the
moment in the director basis (also called {\it torsional moment}), is
constant along the rod, consequently the twist density $k_3$ is
inversely proportional to the twisting coefficient $C(s)$
\begin{equation}
\label{k3}
k_3(s)=\frac{M_3}{C(s)} \; .
\end{equation}

In order to look for circular (ring) and helical solutions of the
Eqs.~(\ref{a}--\ref{f}) the components of the twist vector
$\mathbf{k}$ are expressed as follows:
\begin{eqnarray}
k_1&=&k_F(s)\sin\xi \; , \label{hel_k1} \\
k_2&=&k_F(s)\cos\xi \; , \label{hel_k2} \\
k_3&=&\xi'+\tau_F(s) \; , \label{hel_k3}
\end{eqnarray}
where $k_F(s)$ and $\tau_F(s)$ are the curvature and torsion,
respectively, of the rod centerline, and $\xi$ is given by
Eq.~(\ref{def_xi}). 

Substituting Eqs. (\ref{hel_k1}--\ref{hel_k3}) in
Eqs. (\ref{a}--\ref{f}), extracting $f_1$ and $f_2$ from Eqs. (\ref{e})
and (\ref{d}), respectively, differentiating them with respect to $s$,
and substituting in Eqs. (\ref{a}), (\ref{b}) and (\ref{c}), gives the
following set of nonlinear differential equations:
\begin{eqnarray}
[M_3\,k_F-B\,k_F\,\tau_F]'-(B\,k_F)'\,\tau_F
=0 \; ,  \label{dif_1} \\
(B\,k_F)''+k_F\,\tau_F[M_3-B\,\tau_F]
-f_3\,k_F=0 \; , \label{dif_2}  \\
(B\,k_F)'\,k_F+f'_3=0 \; , \label{dif_3}  
\end{eqnarray}
where we have omitted the dependence on $s$ to simplify the
notation. Appendix A presents the details of the derivation of
Eqs. (\ref{dif_1}--\ref{dif_3}).

The Eqs. (\ref{dif_1}--\ref{dif_3}) for the curvature, $k_F$, and
torsion, $\tau_F$, do not depend on the twisting coefficient,
$C(s)$. Therefore, the centerline of an inhomogeneous rod does not
depend on the twisting coefficient unlike the case of homogeneous rods
(see, for example, Eqs. (13) and (14) of Ref.~\cite{neuk}).

Langer and Singer~\cite{langer} have obtained a set of first-order
ordinary differential equations for the curvature and torsion of the
centerline of a homogeneous rod that contains terms proportional to
$k_F^2$ and $\tau^2_F$. The Eqs. (\ref{dif_1}--\ref{dif_3}) have the
advantage of involving only terms linear in $k_F$ and $\tau_F$.

\section{Planar solutions of inhomogeneous rods}

Planar solutions are obtained by setting $\tau_F=0$ in
Eqs. (\ref{dif_1}-\ref{dif_3}).  The result is:
\begin{eqnarray}
M_3\,k_F'=0 \; ,  \label{p1} \\
(B\,k_F)''-f_3\,k_F=0 \; , \label{p2}  \\
(B\,k_F)'\,k_F+f'_3=0 \; , \label{p3}  
\end{eqnarray}

Eq. (\ref{p1}) implies two different sets of solutions: $M_3\neq0$
(twisted rods) and $M_3=0$ (non-twisted rods). Each case will be
analysed separately.

\subsection{Twisted rod: $M_3\neq0$}

As ($M_3\neq0$), Eq. (\ref{p1}) leads to $k_F(s)=$ Constant
$\equiv K$, which represents the circular solution (also known as {\it
planar ring solution}). Eqs. (\ref{p2}) and (\ref{p3}) must be
also satisfied, and their combination  gives the following differential
equation for the bending stiffness $B(s)$:
\begin{equation}
\label{eqBaux}
\frac{d}{ds}\left[B''+K^{2}B\right]=0 \; ,
\end{equation}
or
\begin{equation}
\label{eqB}
B''+K^{2}B=\mbox{Constant }\equiv C_0 \; ,
\end{equation}
where we have set $k_F=K$, and $C_0$ is a constant of
integration. Therefore, the twisted ($M_3\neq0$) ring solution of an
inhomogeneous rod only exists if its bending stiffness satisfies the
differential equation (\ref{eqB}) (forced harmonic oscillator).  The
bending stiffness (solution of (\ref{eqB})) can be written in the
form:
\begin{equation}
\label{ring}
B(s)=A_0\cos(K\,s)+B_0\sin(K\,s)+ C_0/K^{2} \; ,
\end{equation}
with $A_0$ and $B_0$ arbitrary constants.

Therefore, only an intrinsically straight inhomogeneous twisted rod
($M_3\neq0$) with the bending stiffness given by Eq. (\ref{ring}), can
display a planar ring configuration.

\subsection{Non-twisted rod: $M_3=0$}

In this case, Eq. (\ref{p1}) is automatically satisfied so the
curvature $k_F(s)$ does not have to be constant.

From Eq. ({\ref{p2}), we can obtain the following relation for
$f_3(s)$:
\begin{equation}
\label{f3}
f_3(s)=\frac{(Bk_F)''}{k_F} \; .
\end{equation}
Differentiating $f_3$ with respect to $s$ and substituting in
Eq. (\ref{p3}), we obtain the following non-linear differential
equation for the curvature $k_F(s)$:
\begin{equation}
\label{NLeq}
k_F(Bk_F)'''-k_F'(Bk_F)''+k_{F}^{3}(Bk_F)'=0 \; .
\end{equation}

If $k_F=$ Constant $\equiv K$ (the planar ring solution) the above
equation gives the following differential equation for $B$:
\begin{equation}
\label{eqB2}
\frac{d}{ds}\left[B''+K^{2}B\right]=0 \; ,
\end{equation}
which is the same Eq. (\ref{eqBaux}) and, therefore, gives the same
condition for $B(s)$ shown in Eq. (\ref{ring}). In this case, $f_3(s)$
is given by:
\begin{equation}
\label{f3per}
f_3(s)=-K^2(A_0\cos Ks+B_0\sin Ks) \; .
\end{equation}

So, the non-twisted planar ring ($k_F=$ Constant) is also an equilibrium
solution of the static Kirchhoff equations if the bending stiffness of
the inhomogeneous rod is given by Eq. (\ref{ring}).

Another solution of Eq. (\ref{NLeq}) for the curvature $k_F(s)$ comes
from making $(Bk_F)'=0$, which gives
\begin{equation}
\label{solk_F}
k_F(s)=\frac{\mbox{Constant}}{B(s)} \; ,
\end{equation}
and, from Eq. (\ref{f3}), $f_3(s)=0$.

It is a very interesting solution because it relates the curvature of
the rod centerline to the bending stiffness. Also, the component of
the force tangent to the rod centerline, $f_3$, is null for the set of
solutions satisfying Eq. (\ref{solk_F}). In the particular case where
$B(s)$ is given by Eq. (\ref{ring}), we can have two possible planar
solutions: the planar ring $k_F(s)=K$, and the solution for which
$k_F(s)$ is given by Eq. (\ref{solk_F}).

A {\it kink} is characterized by a piece of a filament having high
curvature $k_F$ (or small radius of curvature, $R=1/k_F$). The
Kirchhoff approach is not appropriate for modelling kinks in rods
because in the derivation of the Kirchhoff equations it is assumed
that the rod centerline radius of curvature is much larger than the
cross-section radius~\cite{dill}. However, Haijun and
Zhong-can~\cite{hai} used the Kirchhoff rod model to show that
variations in the curvature of a closed rod may lead to instabilities
responsible for the kink formation. Within the limits of validity of
the Kirchhoff model, Eq. (\ref{solk_F}) shows that pieces of a rod
with small bending coefficient may lead to equilibrium configurations
with large curvature. Therefore, our results can help explaining the
onset of formation of kinks, since variations in the curvature of a
filament can be originated by variations in the bending stiffness
along it (Eq. (\ref{solk_F})). In Section V we comment on an
application of this result to small closed DNAs. Figure~\ref{fig1}
illustrates a planar equilibrium solution given by Eq. (\ref{solk_F})
in the case of a rod with periodic variation of its bending stiffness.

The straight rod ($k_1=0$, $k_2=0$) is a trivial solution of the
static Kirchhoff equations, and is a particular case of planar
solutions. In the inhomogeneous case, the twist density, $k_3$, of the
straight rod is not constant. According to Eq. (\ref{k3}), for a given
external torque, $M_3$, the twist is large for pieces of the rod where
the twisting coefficient, $C$, is small. This result could help in
determining the regions where plastic deformations or ruptures start
to occur in rods subjected to large stresses.

\begin{figure}[ht]
  \begin{center}
  \includegraphics[height=50mm,width=70mm,clip]{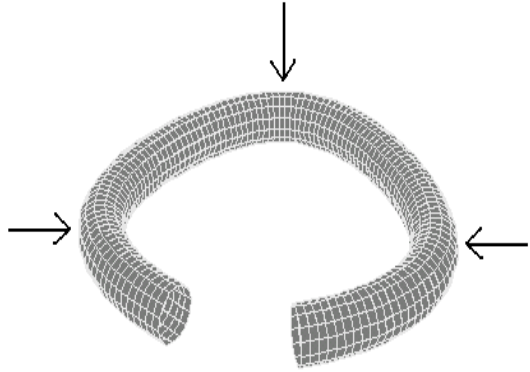}
  \end{center} 
  \caption{A non twisted planar solution of the Kirchhoff equations
  for a rod with periodic variation of the bending stiffness. The
  arrows point to the regions where the curvature is increased. }
\label{fig1}
\end{figure}

\section{Helical solutions of inhomogeneous rods}

In this section, we are considering the equilibrium solutions for the
rod centerline in which $k_F\neq0$ and $\tau_F\neq0$. In order to find
helical solutions for the static Kirchhoff equations, we apply the
{\it Lancret's theorem} to the equations (\ref{dif_1}--\ref{dif_3}).
We first write the {\it Lancret's theorem} in the form:
\begin{equation}
\label{lan2}
k_F(s)=\beta\,\tau_F(s) \; ,
\end{equation}
where $\beta\neq0$ is a constant. Substituting Eq. (\ref{lan2}) in
Eq. (\ref{dif_1}) we obtain
\begin{equation}
\label{eq_B1}
\tau_F'\,(M_3-B\,\tau_F)-2\,\tau_F\,(B\,\tau_F)'=0 \; .
\end{equation}
Substituting Eq. (\ref{lan2}) in Eq. (\ref{dif_2}) and extracting
$f_3$, we obtain
\begin{equation}
\label{f3_2}
f_3=\frac{(B\,\tau_F)''}{\tau_F}+\tau_F\,(M_3-B\,\tau_F) \; .
\end{equation}
Differentiating $f_3$ with respect to $s$ and substituting in
Eq. (\ref{dif_3}) we obtain the following differential equation for
$\tau_F(s)$:
\begin{equation}
\label{eq_B2}
\frac{(B\,\tau_F)'''}{\tau_F}-\frac{(B\,\tau_F)''\tau_F'}{\tau_F^2}+
(\beta^{2}+1)\,\tau_F\,(B\,\tau_F)'=0 \; ,
\end{equation}
where Eq. (\ref{eq_B1}) was used to simplify the above equation. One
immediate solution for this differential equation is
\begin{equation}
\label{sol1}
(B\,\tau_F)'=0 \; ,
\end{equation}
that substituted in Eq. (\ref{eq_B1}) gives
\begin{equation}
\label{aux5}
\tau_F'\,(M_3-B\tau_F)=0 \; .
\end{equation}
For non-constant $\tau_F$, the Eq. (\ref{aux5}) gives the following
solution for $\tau_F$:
\begin{equation}
\label{tf_1}
\tau_F(s)=\frac{M_3}{B(s)} \; .
\end{equation}
If $M_3=0$, the only solution satisfying the {\it Lancret theorem} is
$\tau_F(s)=0$ which is the planar solution given in the previous
section. Substituting Eqs. (\ref{sol1}) and (\ref{tf_1}) in
Eq. (\ref{f3_2}) we obtain that
\begin{equation}
\label{f3Zero}
f_3(s)=0 \; .
\end{equation}
Substituting Eq. (\ref{tf_1}) in (\ref{lan2}) we obtain:
\begin{equation}
\label{kf_1}
k_F(s)=\frac{\beta M_3}{B(s)}  \; .
\end{equation}

Substituting Eq. (\ref{def_xi}) in Eq. (\ref{F}), the force ${\bf{F}}$
becomes
\begin{equation}
\label{F_frenet}
{\bf{F}}=(f_1\cos\xi-f_2\sin\xi)\,{\bf{n}}+
(f_1\sin\xi+f_2\cos\xi)\,{\bf{b}}+f_3\,{\bf{t}} \; ,
\end{equation}
where $\{{\bf{n}},{\bf{b}},{\bf{t}}\}$ is the Frenet basis. Using the
Eqs. (\ref{f1r}) and (\ref{f2r}) for $f_1$ and $f_2$ (Appendix A), we
obtain
\begin{equation}
\label{F2}
{\bf{F}}=-(B\,k_F)'\,{\bf{n}}+k_F\,[M_3-B\,\tau_F]\,{\bf{b}}+
f_3\,{\bf{t}} \; ,
\end{equation}
where in the inhomogeneous case $f_3$ must satisfy the
Eq.~(\ref{dif_3}).

Substituting Eqs. (\ref{tf_1}--\ref{kf_1}) in the Eq. (\ref{F2}), it
follows ${\bf{F}}=0$. Therefore, the helical solutions satisfying
(\ref{sol1}) are {\it free standing}.

Now, we show that a {\it circular helix} cannot be a solution of the
static Kirchhoff equations for an intrinsically straight rod with
non-constant bending stiffness. A circular helix has $k_F=$ Constant
$\neq0$, and $\tau_F=$ Constant $\neq0$, in which case
Eq. (\ref{dif_1}) gives:
\begin{equation}
\label{Blinha}
2\,k_F\,\tau_F\,B'=0 \; .
\end{equation}
Since $k_F\neq0$ and $\tau_F\neq0$, Eq. (\ref{Blinha}) will be
satisfied only if $B'=0$, implying constant bending stiffness.
Therefore, it is not possible to have a {\it circular helix} as a
solution for a rod having non-constant bending coefficient.

The solution for the curvature $k_F$, Eq. (\ref{kf_1}), and the
torsion $\tau_F$, Eq. (\ref{tf_1}), can be used to obtain the unit
vectors of the {\it Frenet} frame $\{
{\mathbf{n}},{\mathbf{b}},{\mathbf{t}} \}$ through the {\it
Frenet-Serret} equations:
\begin{eqnarray}
{\mathbf{t}}'&=&k_F\,{\mathbf{n}} \; \; , \label{sf1} \\
{\mathbf{n}}'&=&-k_F\,{\mathbf{t}}+\tau_F\,{\mathbf{b}} \; \; , \label{sf2} \\
{\mathbf{b}}'&=&-\tau_f\,{\mathbf{n}} \; \; . \label{sf3}
\end{eqnarray}
By choosing the $z$-direction of the fixed cartesian basis as the
direction of the helical axis, we can integrate ${\mathbf{t}}$ in
order to obtain the three-dimensional configuration of the centerline
of the rod.

Figure~\ref{fig2} displays the helical solution of the static Kirchhoff
equations for rods with bending coefficients given by
\begin{eqnarray}
\mbox{Fig \ref{fig2}a:}\;\;\;B_{a}(s)&=&1 \; , \label{r1} \\
\mbox{Fig \ref{fig2}b:}\;\;\;B_{b}(s)&=&1+0.007\,s \; \; , \label{r2} \\
\mbox{Fig \ref{fig2}c:}\;\;\;B_{c}(s)&=&1+0.1\,\sin(0.04s+2) \; \; . \label{r3}
\end{eqnarray}
The case of constant bending (Eq. (\ref{r1})) produces the well known
{\it circular helix} displayed in Fig. \ref{fig2}a, while
Figs.~\ref{fig2}b--\ref{fig2}c display other helical structures
obtained for rods with non-constant bending coefficients
(Eqs. (\ref{r2}--\ref{r3})). These structures are clearly not circular
helices but as they satisfy the {\it Lancret theorem},
Eq. (\ref{lan2}), they are called {\it generalized helices}.

\begin{figure}[ht]
  \begin{center}
  \includegraphics[height=44mm,width=20mm,clip]{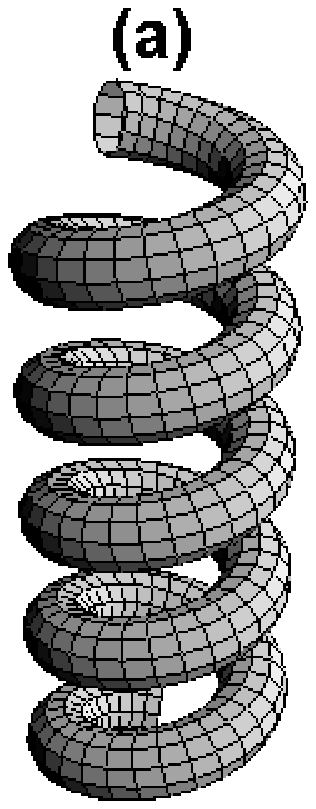}
  \includegraphics[height=40mm,width=18mm,clip]{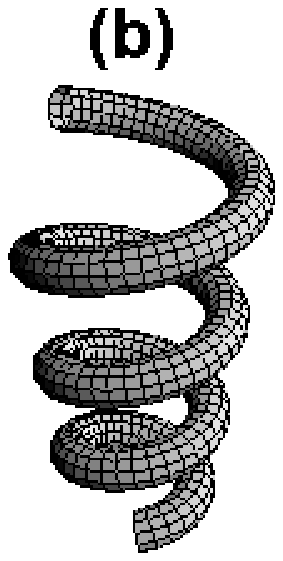}
  \includegraphics[height=44mm,width=22mm,clip]{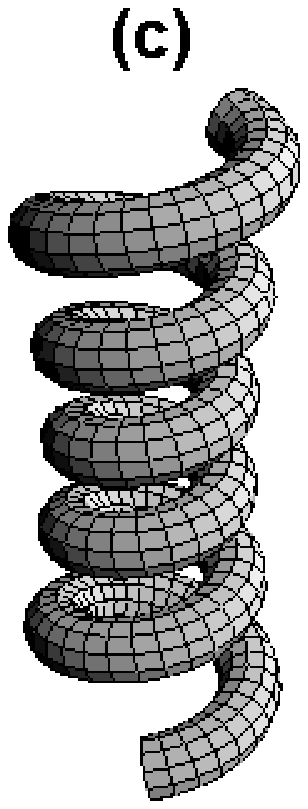}
  \end{center} 
  \caption{Helical solutions of the Kirchhoff equations using the {\it
  Lancret's Theorem}. (a) circular helix solution valid for a rod with 
  constant bending coefficient $B_a=1$ (\ref{r1}); (b) and (c) are 
  generalized helices for inhomogeneous rod with bending coefficient
  given by Eqs. (\ref{r2}) and (\ref{r3}), respectively. The parameters, 
  in scaled units, are $M_3=0.05$, and the total length of the rod is 
  $L=130$. $k_F(0)=0.24$ for the helical solutions displayed in panels 
  (a) and (b), and $k_F(0)=0.22$ for panel (c). }  
\label{fig2}
\end{figure} 

The tridimensional helical configurations displayed in Fig.~\ref{fig2}
were obtained by integrating the {\it Frenet-Serret} equations
(\ref{sf1}-\ref{sf3}) using the following initial conditions for the
Frenet frame: ${\mathbf{t}}(s=0)=(0,\sin\alpha,\cos\alpha)$,
${\mathbf{n}}(s=0)=$~$(-1,0,0)$ and
${\mathbf{b}}(s=0)=$~$(0,-\cos\alpha,\sin\alpha)$, where $\alpha$
represents the angle between the tangent vector ${\mathbf{t}}$ and the
helical axis. This choice ensures that the $z$-axis is parallel to the
direction of the helical axis. The centerline of the helical rod is a
space curve ${\mathbf{x}}(s)=(x(s),y(s),z(s))$ that is obtained by
integration of the tangent vector ${\mathbf{t}}(s)$. We have placed
the initial position of the rod at $x(0)=1/k_1(0)$, $y(0)=0$ and
$z(0)=0$ (in scaled units), where $k_1(0)$ is the curvature of the
projection of the helical centerline onto the plane perpendicular to
the helical axis. It is possible to show~\cite{dirk} that
\begin{equation}
\label{X0}
x(0)=\frac{1}{k_1(0)}=\frac{k_F(0)\,B^2(0)}{M_3^2+k_F^2(0)B^2(0)} \; .
\end{equation}
$k_F(0)$ and $M_3$ are free parameters that have been chosen so that
the helical solutions displayed in Fig.~\ref{fig2} have the same angle
$\alpha$. The parameters $k_F(0)=0.24$ and $M_3=0.05$ give
$x(0)\simeq4$ for the helical solutions displayed in Figs.~\ref{fig2}a
and \ref{fig2}b, and the parameters $k_F(0)=0.22$ and $M_3=0.05$ give
$x(0)\simeq4.36$ for the helical solution displayed in the
Fig.~\ref{fig2}c.

\section{Conclusions}

The existence of circular (planar ring) and helical configurations for
a rod with non-constant bending and twisting stiffnesses has been
investigated within the framework of the Kirchhoff rod model. We are
interested in finding analytical solutions, and since the static
Kirchhoff equations consist of nine non-linear PDEs, we derived from
them a set of differential equations (Eqs.~(\ref{dif_1}--\ref{dif_3}))
for the curvature, $k_F$, and torsion, $\tau_F$, of the rod
centerline, and then applied the geometric conditions to obtain
circular and helical solutions.

We have shown that the planar ring ($k_F=$ Constant) is an equilibrium
solution of the Kirchhoff equations for an inhomogeneous rod which
possesses the bending stiffness, $B(s)$, satisfying Eq. (\ref{ring}),
and $f_3(s)$ satisfying Eq. (\ref{f3per}). The planar ring is the only
planar solution for twisted rods ($M_3\neq0$). Non-twisted rods
($M_3=0$) may present other types of planar solutions with $k_F(s)$
(non-constant) satisfying Eq. (\ref{NLeq}).

We also have shown that the circular helix is a solution of the static
Kirchhoff equations only if the rod bending coefficient $B=$ Constant
(see Eq. (\ref{Blinha})). An inhomogeneous rod has $B\neq$ Constant,
therefore it cannot have a circular helix as equilibrium
solution. Nevertheless, it can exhibit generalized helical structures
that satisfy the {\it Lancret's theorem}, Eq. (\ref{lan2}).  In the
Kirchhoff model, these generalized helical solutions can be obtained
solving Eq. (\ref{eq_B2}) for $\tau_F(s)$. Figure~\ref{fig2} displays
three examples of helical structures that satisfy the static Kirchhoff
equations.

May an intrinsically straight homogeneous elastic rod present other
types of generalized helical equilibrium solutions~? A homogeneous rod
has $C$ constant, implying $k_3=$ Constant (from Eq. (\ref{f})). It
has been proved that $\xi'=0$ for a helical solution of a homogeneous
rod (see reference~\cite{tyler}), and Eq. (\ref{hel_k3}) shows that
the torsion $\tau_F=k_3=$ Constant. In order to satisfy the {\it
Lancret's theorem} (Eq. (\ref{lan2})), the curvature $k_F$ of this
helical solution must also be constant. Since the helical structure
whose curvature, $k_F$, and torsion, $\tau_F$, are constant, is a
circular helix, {\it the only type of helical solution for an
intrinsically straight homogeneous rod is the circular helix}.

Some motivations for this work are related to defects~\cite{kronert}
and distortions~\cite{geetha} in biological molecules. Also, the
stiffness of the DNA molecule has been proved to be
sequence-dependent~\cite{tamar}. These defects, distortions and
sequence-dependent elastic properties could be modeled as
inhomogeneities along a continuous elastic rod.

Haijun and Zhong-can~\cite{hai} have used the Goriely and Tabor
stability analysis method~\cite{gori1,goriely} to explain the onset of
instability in the phenomenon of kink formation in small closed DNA
molecules reported by Han {\it et al}~\cite{han1,han2}. In the Han
{\it et al} experiment~\cite{han2}, the DNA rings are immersed in a
solution containing Zn$^{2+}$ or Mg$^{2+}$ ions. They found that DNA
rings become kinked when the concentration of Zn$^{2+}$ ions is above
a critical value, with or without the presence of Mg$^{2+}$ which
alone did not produce the kinking phenomenon. Haijun and Zhong-can
proposed that the binding of Zn$^{2+}$ ions to DNA basepairs enhance
the intrinsic curvature of the DNA destabilizing it. The Mg$^{2+}$
ion, by binding to the phosphate backbone, did not produce intrinsic 
curvature in the DNA molecules. Nevertheless, Haijun and Zhong-can
theory did not explain how the binding of Zn$^{2+}$ ions lead to a
variation of the curvature of the closed DNA. Also, the asymmetries
and positions of the kinks in the DNA rings have not been explained.

Our approach can give a new insight to the phenomenon of kink
formation of these closed DNAs. Due to the binding of the Zn$^{2+}$
ion to a basepair, the bending stiffness of the DNA molecule should
decrease locally, leading to a local increase of curvature. On the
other hand, the binding of the Mg$^{2+}$ ion to the phosphate backbone
cannot change the stiffness of the molecule since the phosphate
backbone is a very rigid structure. Our model supports this proposal
since the equilibrium solutions represented by Eq.~(\ref{solk_F}) show
that the curvature is inversely proportional to the bending
stiffness. Also Eq. (\ref{solk_F}) is only valid for a non-twisted rod
which is the case of the DNA rings reported in Ref.~\cite{han1,han2}.

Haijun and Zhong-can have shown that closed DNA with curvature above a
critical value never forms kinks in presence of Zn$^{2+}$ or Mg$^{2+}$
ions. It explains why Han {\it et al}~\cite{han2} frequently observed
kinks in closed DNAs of 168 basepairs (bps) long, while they rarely
observed kinks in closed DNAs of 126-bps long, independent of the
presence of Zn$^{2+}$ ions in the solution containing the DNA
rings. However, Haijun and Zhong-can approach cannot explain the
approximate elliptical shapes of the 126-bps circles observed by Han
{\it et al}~\cite{han2}. Our results suggest that the non-circular
shape of the 126-bps closed DNAs is a consequence of the binding of
Zn$^{2+}$ ions to the basepairs that decreases locally the stiffness
of the DNA, thus increasing the curvature.

It has been proposed~\cite{schell,crick,sobell} that the kinking
phenomenon is an important mechanism for wrapping DNA around
nucleosomal proteins. In 1997, Luger {\it et al}~\cite{luger} reported
the X-ray crystalline structure of the nucleosomal protein,
demonstrating that the DNA is not uniformly bent around nucleosome,
but presents minimum and maximum curvatures at different positions. It
is well known that the stiffness of DNA is sequence-dependent. Our
results show that the variation of the bending stiffness of the DNA
can give rise to these minimum and maximum curvatures.

\section*{Acknowledgements}
This work was partially supported by the Brazilian agencies FAPESP,
CNPq and CAPES. The authors would like to thank Prof. Manfredo do
Carmo for valuable informations about the Lancret's theorem.

\begin{appendix}

\section{Appendix: The differential equations for the curvature and 
torsion}

Here, we shall derive the
Eqs. (\ref{dif_1}--\ref{dif_3}). Substitution of
Eqs. (\ref{hel_k1}--\ref{hel_k3}) into Eqs. (\ref{a}--\ref{f}) gives:
\begin{eqnarray}
\label{es1_r}
f'_1-f_2\,(\xi'+\tau_F)+f_3\, k_F\cos\xi=0 \; , \label{ar}  \\
f'_2+f_1\,(\xi'+\tau_F)-f_3\, k_F\sin\xi=0 \; , \label{br}  \\
f'_3-f_1\, k_F\cos\xi+f_2\, k_F\sin\xi=0 \; , \label{cr}  \\
(B(s)\, k_F\sin\xi)'+(C(s)-B(s))\, k_F\cos\xi\,(\xi'+\tau_F)-f_2=0 \; , 
\label{dr} \\
(B(s)\, k_F\cos\xi)'-(C(s)-B(s))\, k_F\sin\xi\,(\xi'+\tau_F)+f_1=0 \; , 
\label{er} \\
(C(s)\ (\xi'+\tau_F))'=0 \; . \label{fr} 
\end{eqnarray}

First, we extract $f_1$ and $f_2$ from Eqs. (\ref{er}) and (\ref{dr}),
respectively:
\begin{eqnarray}
\label{fs}
&&f_1=-(B(s)\,k_F)'\cos\xi+[M_3\,k_F-B(s)\,k_F\,\tau_F]\sin\xi
\; , \label{f1r} \\
&&f_2=(B(s)\,k_F)'\sin\xi+[M_3\,k_F-B(s)\,k_F\,\tau_F]\cos\xi
\; , \label{f2r}
\end{eqnarray}
where $M_3=C(s)\,(\xi'+\tau_F)$ is the torsional moment of the rod
that is constant by Eq. (\ref{fr}). Differentiating $f_1$ and $f_2$
with respect to $s$, substituting in Eqs. (\ref{ar}) and (\ref{br}),
respectively, and using Eqs. (\ref{f1r}) and (\ref{f2r}), gives the
following equations:
\begin{equation}
\label{eq_A}
\begin{array}{ll}
\left\{-(B(s)\,k_F)''-\tau_F[M_3\,k_F-B(s)\,k_F\,\tau_F]
+f_3\,k_F\right\}\,\cos\xi \, \, \, + \\
\left\{[M_3\,k_F-B(s)\,k_F\,\tau_F]'
-\tau_F(B(s)\,k_F)'\right\}\, \sin\xi = 0 \; ,
\end{array}
\end{equation}
\begin{equation}
\label{eq_B}
\begin{array}{ll}
\left\{(B(s)\,k_F)''+\tau_F[M_3\,k_F-B(s)\,k_F\,\tau_F]
-f_3\,k_F\right\}\,\sin\xi \, \, \, + \\
\left\{[M_3\,k_F-B(s)\,k_F\,\tau_F]'
-\tau_F(B(s)\,k_F)'\right\}\, \cos\xi = 0 \; .
\end{array}
\end{equation}
Multiplying Eq. (\ref{eq_A}) (Eq. (\ref{eq_B})) by $\sin\xi$
($\cos\xi$) and then adding the resulting equations, we obtain the
Eq. (\ref{dif_1}) for the curvature and torsion:
\begin{equation}
\label{dif1r}
[M_3\,k_F-B\,k_F\,\tau_F]'-(B\,k_F)'\tau_F=0 \; .
\end{equation}
Multiplying Eq. (\ref{eq_A}) (Eq. (\ref{eq_B})) by $-\cos\xi$
($+\sin\xi$) and then adding the resulting equations, we obtain the
Eq. (\ref{dif_2}):
\begin{equation}
\label{dif2r}
(B\,k_F)''+k_F\,\tau_F(M_3-B\tau_F)-f_3\,k_F=0 \; .
\end{equation}
Finally, the Eq. (\ref{dif_3}) is obtained by substituting
Eqs. (\ref{f1r}) and (\ref{f2r}) in Eq. (\ref{cr}):
\begin{equation}
\label{dif3r}
(B\,k_F)'+f_3'=0 \; .
\end{equation}

\end{appendix}


}
\end{doublespace}


\begin{thebibliography}{99}
\bibitem{whip} A whip is a nonhomogenous thread with varying radius of
cross section.
\bibitem{foot2} The references in \cite{dill} can be seen for a 
complete derivation of the Kirchhoff equations.
\end{thebibliography}
\end{document}